\documentclass{article}

     \PassOptionsToPackage{numbers, compress}{natbib}

    \usepackage[preprint]{neurips_2020}

\usepackage[utf8]{inputenc} 
\usepackage[T1]{fontenc}    
\usepackage{hyperref}       
\usepackage{url}            
\usepackage{booktabs}       
\usepackage{amsfonts}       
\usepackage{nicefrac}       
\usepackage{microtype}      

\usepackage[noend]{algpseudocode}
\usepackage{algorithmicx,algorithm}
\usepackage{graphicx}
\usepackage{float} 
\usepackage{subfigure}
\usepackage{bm}

\title{Embodied Self-supervised Learning by Coordinated Sampling and Training}

\author{
  Yifan Sun \\
  Peking University \\
  \texttt{yifan\_sun@pku.edu.cn} \\
  \And
  Xihong Wu \\
  Peking University \\
  \texttt{wxh@cis.pku.edu.cn} \\
}

\begin{document}

\maketitle

\begin{abstract}
  Self-supervised learning can significantly improve the performance of downstream tasks, 
  however, the dimensions of learned representations normally lack explicit physical meanings. 
  In this work, we propose a novel self-supervised approach to solve inverse problems by employing the corresponding
  physical forward process so that the learned representations can have explicit physical meanings.
  The proposed approach works in an analysis-by-synthesis manner to learn an inference network
  by iteratively sampling and training. 
  At the sampling step, given observed data, the inference network is used to approximate the intractable posterior, 
  from which we sample input parameters and feed them to a physical process to generate data in the observational space; 
  At the training step, the same network is optimized with the sampled paired data. 
  We prove the feasibility of the proposed method by tackling the acoustic-to-articulatory inversion problem 
  to infer articulatory information from speech. 
  Given an articulatory synthesizer, an inference model can be trained completely from scratch with random initialization.
  Our experiments demonstrate that the proposed method can converge steadily and the network learns 
  to control the articulatory synthesizer to speak like a human.
  We also demonstrate that trained models can generalize well to unseen speakers or even new languages, 
  and performance can be further improved through self-adaptation. 
\end{abstract}

\section{Introduction}
  In recent years, self-supervised learning has attracted lots of attention, of which the typical 
  pipeline can be described as follows: generate pseudo labels from data itself, perform supervised 
  learning to learn representations, and then transfer the learned representations for downstream 
  tasks. Such methods have achieved great successes in improving downstream tasks performance and reducing 
  the demand of labelled data in many fields such as computer vision \citep{chen2020simple,kolesnikov2019revisiting,xie2016unsupervised}, 
  natural language processing \citep{devlin2019bert}, and speech processing \citep{jiang2019improving}. However, despite the various 
  constraints proposed, the dimensions of learned representations are normally lack of
  explicit physical meanings. Therefore, it is an important but unsolved problem of how to make a data 
  representation generated from self-supervised learning have an explicit physical meaning so that 
  the representation becomes explainable. 

  Given an observed data, there must be an underlying physical process that generates it with certain input.
  If we have access to such a physical forward process, either physical or physically simulated,
  we may design a learning procedure to infer the latent input parameters as a physically meaningful representation of the observed data. 
  Normally, to solve such an inverse problem, the physical process can play a role as the decoder in auto-encoder parlance 
  that generates an output for supervision by taking the inferred parameters as input. 
  However, there are several challenges:
  \begin{itemize}
    \item Non-differentiable physical process. The physical process may be non-differentiable w.r.t. its input, 
    thus auto-encoder like methods cannot be directly applied.
    \item Non-trivial to sample data. Given a physical process, it is natural to consider constructing a synthetic paired 
    training dataset by sampling inputs first and then generating outputs. 
    However, an explicit prior of input parameters is normally difficult to obtain, 
    while random sampling can be computationally infeasible and may end up with meaningless samples.
    \item Generalization issues. Even when priors or samples of input parameters are available, 
    a trained model may suffer the generalization problem when applied to unseen observations.
  \end{itemize}
  To tackle these challenges, we propose a novel analysis-by-synthesis procedure by iteratively sampling and
  training. At the sampling step, given an observed data, a neural network is used to approximate the 
  intractable posterior of input parameters, then parameters are sampled from the posterior
  to generate outputs in the observational space through the physical process; At the training step, the same 
  network is trained on the sampled paired data to predict input parameters from observations. These 
  two steps operate iteratively and boost each other, similar to the guess-try-feedback process in human 
  learning. The entire learning procedure integrates the physical generation process to obtain 
  meaningful parameter estimates in a self-supervised manner. The proposed method can address the 
  above-mentioned three challenges as follows:
  \begin{itemize}
    \item The physical process is only used for generating outputs, without passing gradients through.
    \item No need of priors on input parameters, the model learns from unlabelled data to approximate 
    the posterior of input parameters for efficiently sampling.
    \item The self-supervised nature of the training procedure means a trained model can adapt itself to unseen observations.
  \end{itemize}
 
  We verify the proposed method by tackling the acoustic-to-articulatory inversion problem to extract articulatory 
  kinematics information from speech. We adopt the Tube Resonance Model (TRM) \citep{hill2017low} to simulate the 
  human vocalization mechanism. Given unlabelled reference utterances and starting completely from scratch with random initialization, a 
  network can learn to infer how to control the TRM model to synthesize sound similar 
  to the reference utterances by alternately sampling data and training itself. Experiments show that our proposed 
  algorithm can converge steadily. The synthesized sounds achieve a signal to noise ratio of around 16dB on both single-speaker 
  and multi-speaker datasets. Further experiments show that trained models can generalize well to unseen speakers or even a new language, 
  and performance can be further promoted through self-adaptation. 

\section{Related Work}
\paragraph{Inverse Problem}
  Inverse problems aim to infer causes from observations. Traditionally, analytical methods have been used
  to solve inverse problems such as image super-resolution (SR) \citep{park2003super}, 
  computed tomography (CT) \citep{natterer2001mathematics}. 
  In recent years, data-driven methods especially the deep learning approach has achieved state-of-the-art 
  results \citep{brooks2019unprocessing, liu2019coherent}.
  For those problems with a differentiable or invertible forward process,
  methods have been proposed to solve problems like SR \citep{sonderby2016amortised}.
  Otherwise, the key to solve an inverse problem in a data-driven way is to obtain supervision data.
  When a physical forward process is available,
  paired training data can be synthetically generated by applying the forward process to samples of input parameters 
  (e.g., adding noise to clean images for denoising training).
  However, appropriate priors or samples of the input parameters are not always available, 
  while the parameter space can be high dimensional, making input parameters sampling non-trivial.

  To solve an inverse problem for a given observed data, sampling from the posterior distribution can be efficient. 
  Markov chain Monte Carlo (MCMC) is a widely used sampling method, but it requires an explicit prior 
  and can be computationally infeasible in a high-dimensional space. 
  As for deep sampling methods, \citep{adler2018deep} proposed a method to directly sample from the posterior 
  with a conditional generative adversarial network (CGAN), but this method requires supervised training. 
  Variational auto-encoders (VAEs) \citep{diederik2014auto} directly approximate the posterior of latent variables with a neural network. 
  Similar to that, we use a neural network to approximate the posterior of parameters and sample from it to generate paired training data. 
  However, since the forward process (corresponding to the decoder in VAE) can be non-differentiable, 
  we turn to minimize the reconstruction error of latent input parameters rather than observed data.
\paragraph{Acoustic-to-Articulatory Inversion (AAI)}
  The motor theory \citep{liberman1985motor} indicates that when perceiving speech, except for acoustic features, human also perceive 
  articulatory information. There has been a lot of work to infer articulatory kinematics information from speech. 
  \citep{afshan2015improved, mitra2017joint, chartier2018encoding} train neural networks for acoustic-to-articulatory inversion on simultaneously recorded speech acoustics and articulatory data in a supervised manner. Since articulatory data is often recorded by ElectroMagnetic Articulography (EMA), 
  or Magnetic Resonance Imaging (MRI), massive articulatory recordings can be too expensive to obtain. 
  Besides, trained models are always speaker-dependent. 

  Given an articulatory synthesizer, 
  either mechanical \citep{fukui2009three, yoshikawa2003constructivist} or physically simulated \citep{howard2014learning}, 
  articulatory synthesis takes a time series of articulatory parameters to produce speech signal. 
  Several methods \citep{higashimoto2002speech, asada2016modeling}
  have been proposed to infer proper articulatory parameters to reproduce a given speech by building parameter-sound
  pair datasets first and then training a model with the datasets. These approaches deal with either a single vowel or a single word with two to three syllables, thus, it is possible to build those datasets. \citep{gao2019articulatory} developed a generic method to 
  do copy-synthesis of speech with good quality, but the method is time consuming when applied for inference. 
  In this work, we adopt a simulated vocalization model as synthesizer to reproduce speech with arbitrary 
  length.
\section{Approach}
\subsection{Problem Senario}
  Traditionally, an inverse problem is formulated to solve an equation in the form of $\mathbf{x} = F (\mathbf{z}) + \mathbf{e}$, where 
  $\mathbf{x} \in X$ is the measured data (observation); $\mathbf{z} \in Z$ is the input parameter (latent variable) of the forward operator $F$; both $\mathbf{x}$ and $\mathbf{z}$ can be high dimensional vectors; $F$ describes how measured data is generated from the 
  input parameters in the absence of noise; $\mathbf{e} \in X$ is the observational noise. 
  For clarity, we include the noise model into the forward operator $F$ in the following description. 
  In case that $F$ is non-linear and not differentiable w.r.t. its parameters, it will be difficult to apply analytical methods. 
  We consider solving the inverse problem through learning. 

  Given $N$ $i.i.d.$ supervised training data $\{(\mathbf{z}_i, \mathbf{x}_i)\}_{i=1}^{N}\subset Z \times X$, an inverse problem 
  can be formulated to find an operator\(R_{\hat\theta}:X\rightarrow Z\), where $\hat\theta$ solves:
  \begin{equation}
  \label{eq1}
    \hat\theta:=\mathop{\arg\min}_{\theta}\frac{1}{N}\sum_{i=1}^{N}-logP(\mathbf{z}_i|R_{\theta}(\mathbf{x}_i))
  \end{equation}
  In this work, we aim to solve an unsupervised inverse problem, i.e., given $N$ $i.i.d.$ observed 
  data $\{\mathbf{x}_i\}_{i=1}^{N} \subset X$ and a forward operator $F$, to find an operator 
  \(R_{\hat\theta}:X \rightarrow Z\), where $\hat\theta$ solves: 
  \begin{equation}
  \label{eq2}
    \hat\theta:=\mathop{\arg\min}_{\theta}\frac{1}{N}\sum_{i=1}^{N}[-logP(\mathbf{x}_i|R_{\theta}(\mathbf{x}_i)) + S_{\lambda}(R_{\theta}(\mathbf{x}_i))]
  \end{equation}
  In which $-logP(\mathbf{x}|R_{\theta}(\mathbf{x}))$ is the reconstruction error, while $S_{\lambda}$ is the potential prior constraints on the parameters. If observational noise is additive Gaussian, 
  $-logP(\mathbf{x}|R_{\theta}(\mathbf{x}))\propto ||\mathbf{x} - F(R_{\theta}(\mathbf{x}))||^2$. When the reconstruction error is small enough, it is fair to say $R_{\theta}(\mathbf{x})$ is a solution to the inversion of $\mathbf{x}$, regardless of prior constraints.
\subsection{Method}
  Given a forward operator $F$, in principle, we can sample from the parameter space $Z$, and then apply 
  the forward operator $F$ to obtain paired data for training. However, in many problems, it is non-trivial to obtain 
  appropriate priors while the latent space can be high dimensional, thus random sampling can be 
  computationally infeasible. 
  
  Given an observational data ${\mathbf{x}_i}$, it can be efficient to sample around 
  the corresponding latent variable ${\mathbf{z}_i}$ (which we do not know), that is, sampling from the posterior
   given ${\mathbf{x}_i}$, from a Bayesian perspective. The problem is that the explicit expression of the posterior
  is normally intractable to obtain. However, consider Equation (\ref{eq1}) for supervised learning, 
  $P(\cdot|R_{\theta}(\mathbf{x}_i))$ is actually the posterior distribution given ${\mathbf{x}_i}$. 
  Intuitively, given an observed data ${\mathbf{x}_i}$, suppose we can approximate the posterior distribution well, 
  we can sample ${\mathbf{z}_i}$ from it and then generate training data, which in turn can help 
  optimize the approximation of the posterior distribution in a supervised way.

  From such an intuition, we propose to solve the unsupervised inverse problem with an iteratively sampling 
  and training procedure, as shown in Algorithm 1. 
  At the sampling step, a neural network is used to approximate the posterior of latent variables, we then sample from the posterior 
  and apply $F$ to generate paired supervised data; At the training step, the same network is trained to 
  optimize the posterior distribution approximation with the sampled supervision data. 
  After several iterations, we can obtain a good approximation of the posterior through 
  the trained network, which in turn means the reconstruction error in Equation (\ref{eq2}) of the 
  predicted parameters can be small enough to consider the unsupervised inverse problem solved. 
  When the learning procedure ends, we can run the inference network in a deterministic way to obtain inversion results.
  The proposed approach could be viewed as a joint embedding architecture,
  from a perspective that the reference utterance and the re-synthesized utterance are distorted versions of the same speech 
  sound, the same network used in the sampling step and the training step are twin networks sharing the same parameters, 
  and the training procedure works to adjust the parameters of the networks so that their embeddings (articulatory parameters) come closer.
  Since the distorted synthetic utterances are generated through an embodied process with an articulatory synthesizer,
  we call Algorithm \ref{algo1} as EMbodied Joint EMbedding (EmJEm). 
  
  \begin{algorithm}[t]
    \label{algo1}
    \caption{Embodied Self-supervised Learning}
    \hspace*{0.02in} {\bf Input:} 
    Observation dataset \(\{\mathbf{x}_{i}\}_{i=1}^{N}\), traininng set \(\Phi=\emptyset\), forward operator \(F\),
    inference neural network \(R_{\theta}\), sample number per datapoint \(L\), iteration limit \(T\), epoch number \(E\), batch size \(M\) \\
    \hspace*{0.02in} {\bf Output:} 
    optimized parameters \(\theta\)
    \begin{algorithmic}[1]
      \State \(\theta\leftarrow\) Random initialize parameters 
      \For{iteration t = 1 to T}
        \State *** sampling step start ***
        \State \(\theta^t \leftarrow \theta\) (Set the latest \(\theta\) for calculating posterior in iteration t)
        \State \(\Phi_t\leftarrow \emptyset\) (Initialize an empty set of sampled paired data for current iteration)
        \For{each \(\mathbf{x}_{i}\) in \(\{\mathbf{x}_{i}\}_{i=1}^{N}\)}
          \State \(Q_{t,i}(\mathbf{z})\leftarrow P(\mathbf{z}|R_{{\theta}^t}(\mathbf{x}_i))\) (Approximate posterior)
          \State \(\{\mathbf{z}_{t,i}^{(l)}\}_{l=1}^L\leftarrow\) Draw L samples from \(Q_{t,i}(\mathbf{z})\)
          \State \(\{(\mathbf{z}_{t,i}^{(l)}, \mathbf{x}_{t,i}^{(l)})\}_{l=1}^L\leftarrow\) Generate paired data, \(\mathbf{x}_{t,i}^{(l)}=F(\mathbf{z}_{t,i}^{(l)})\)
          \State \(\Phi_t \leftarrow \Phi_t\bigcup\{(\mathbf{z}_{t,i}^{(l)}, \mathbf{x}_{t,i}^{(l)})\}_{l=1}^L\)(Update current sampling set)
        \EndFor
        \State \(\Phi \leftarrow \) Update training set \(\Phi\) with newly sampled paired data set \({\Phi_t}\)
        \State *** training step start *** 
        \For{epoch e = 1 to E}
          \For{number of batches in a training epoch}
          \State\(\{(\mathbf{z}^{(m)}, \mathbf{x}^{(m)})\}_{m=1}^M \leftarrow\) Random minibatch of M samples from training set \(\Phi\)
          \State\(\theta \leftarrow \nabla_\theta \frac{1}{M}\sum_{m=1}^{M}-logP(\mathbf{z}^{(m)}|R_{\theta}(\mathbf{x}^{(m)}))\) (Update parameters using gradients)
          \EndFor
        \EndFor
      \EndFor
      \State \Return \(\theta \)
    \end{algorithmic}
    \end{algorithm}

  In practice, if $\mathbf{z}$ is continuous, it is common to describe the posterior with the multivariate 
  Gaussian: \(P(\mathbf{z}|R_{\theta}(\mathbf{x})) = \mathcal{N}(\mathbf{z}; R_{\theta}(\mathbf{x}),\sigma^{2}\mathbf{I})\), 
  in which $\sigma$ is a superparameter.
  We noticed in experiments that if given enough observed data, the sample number per 
  datapoint $L$ can be set to 1, so that the deterministic output of the inference model $R_{\theta}(\mathbf{x})$ can be 
  directly adopted as a sample. 
  Otherwise, sampling several latent variables and weighting the samples as in \citep{burda2016importance} may help.

  There are many options to update the training set $\Phi$. The simplest way is to replace the training 
  set with the latest sampled dataset, i.e. $\Phi=\Phi_t$; to fully utilize historically sampled 
  data and stablize the learning procedure, we can drop part of the historical data at a fixed percentage in each iteration, 
  then merge the left historically sampled data with the latest sampled set, as shown in Algorithm 2.

  \begin{algorithm}
    \label{algo2}
    \caption{Update Training Set with Historical data} 
    \hspace*{0.02in} {\bf Input:} 
    Current training set \(\Phi\), newly sampled training set \(\Phi_t\), drop rate \(\gamma\)\\
    \hspace*{0.02in} {\bf Output:} 
    Updated training set \(\Phi\)
    \begin{algorithmic}[1]
      \State \(\Phi \leftarrow\) Randomly drop samples in \(\Phi\) with a ratio given by \(\gamma\)
      \State \(\Phi \leftarrow \Phi\bigcup\Phi_t\)
      \State \Return \(\Phi\)
    \end{algorithmic}
  \end{algorithm}

\section{Experimental Setup}
  We verify the proposed method by tackling the acoustic-to-articulatory inversion problem. 
  In this section, we introduce the adopted articulatory synthesizer, model architecture and training procedure. 
  Training setup and evaluation metrics are also provided.
\subsection{Articulatory Synthesizer}
  We adopt the Tube Resonance Model (TRM) \citep{hill2017low}, which simulates the propagation of sound waves through 
  a tube by waveguide techniques, as our articulatory synthesizer. Composed of a vocal tract 
  with 8 segments and a nasal cavity with 5 segments, the TRM accepts 26-dimensional utterance-rate parameters 
  and 16-dimensional time varying control-rate parameters as input to synthesize sound. Utterance-rate parameters specify the 
  global state of the tube, such as tube length, glottal pulse, breathness. Control-rate parameters 
  dynamically control the tube to produce time varying sounds by changing diameters of segments and velum, 
  setting micro intonations, and inserting fricatives or aspiration as needed.
  For details of the TRM model, please refer to \citep{hill2017low}. Hereinafter, we refer to the utterance-rate
  parameters and control-rate parameters as articulatory parameters.
  \begin{figure}
    \centering  
    \subfigure[Sampling and training steps in an iteration]{
    \label{Fig.whole}
    \includegraphics[width=0.49\textwidth]{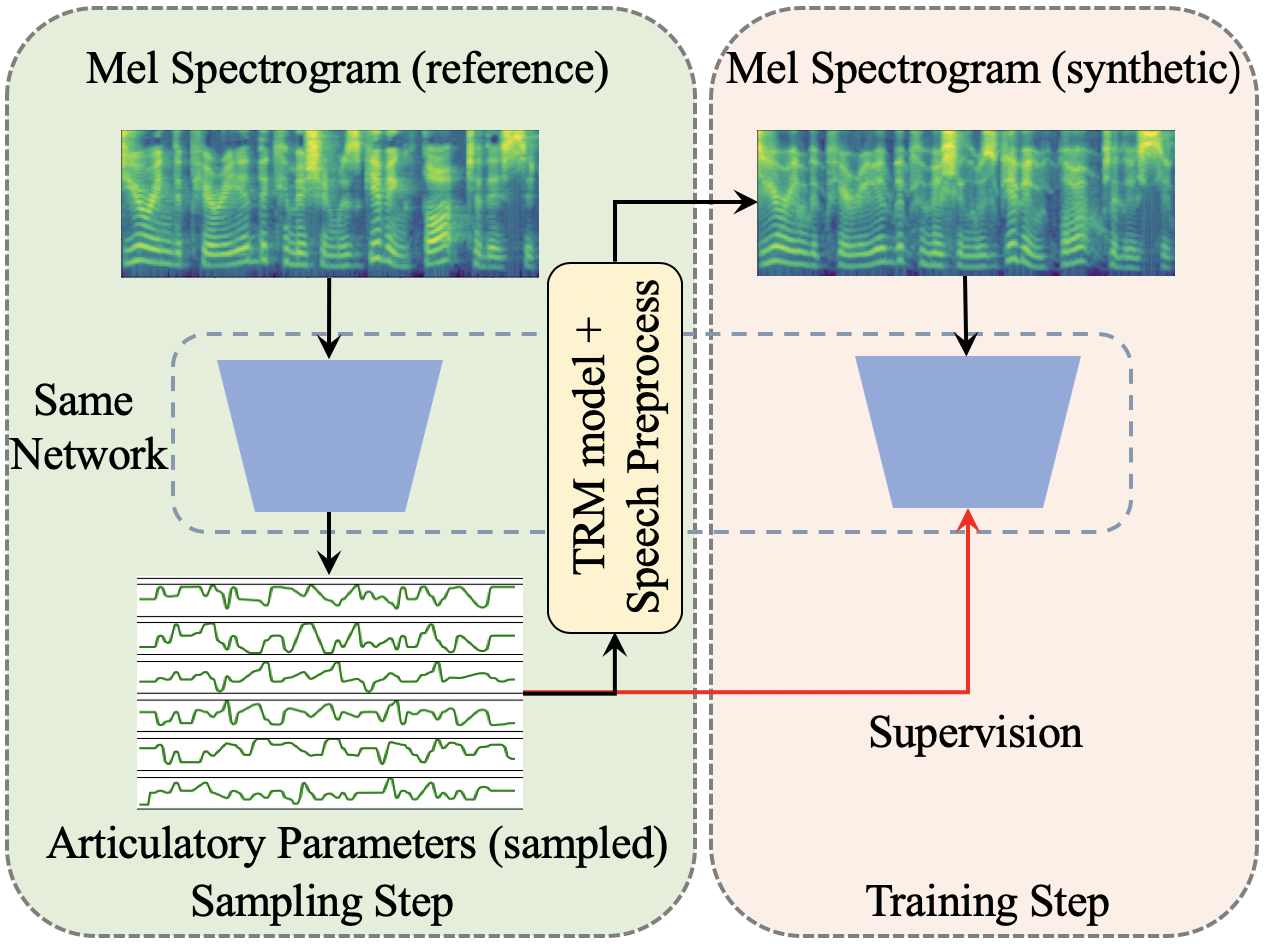}}
    \subfigure[Articulatory parameters inference model]{
    \label{Fig.architecture}
    \includegraphics[width=0.49\textwidth]{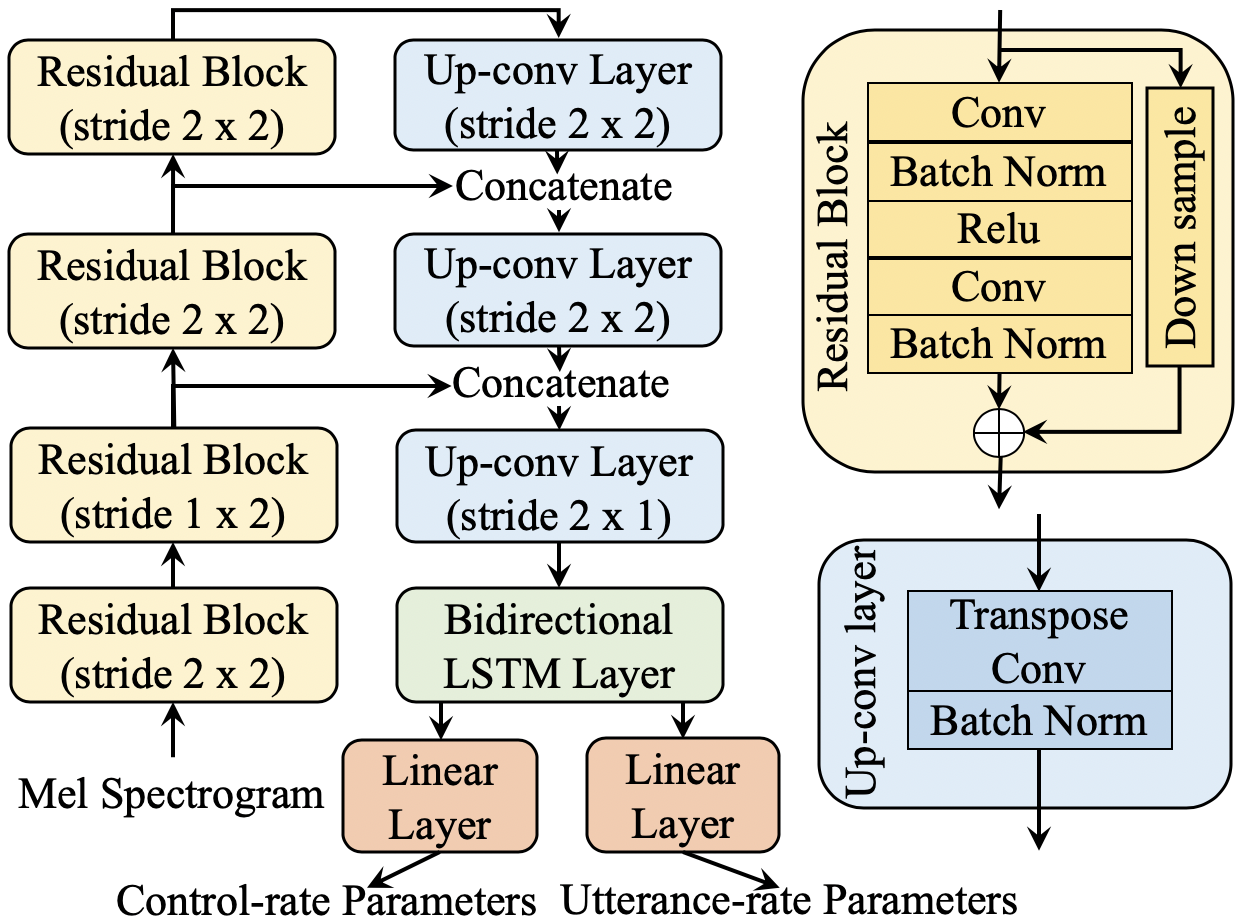}}
    \caption{Overview of an iteration (a) and the model architecture (b). 
    (a) At the sampling step, output of the network is directly adopted as a sampling result; 
    (b) Strides are specified to conserve sequence lengths, 
     when input a mel-spectrogram with shape $N\times80$ (N denotes the frame number), 
     the output of the last up-conv layer will have shape $N\times20$, 
     ignoring dimensions of batchsize and channels.}
    \label{Fig.main}
  \end{figure}
\subsection{Learning Procedure and Model Configuration}
  Given speech data and the TRM model, we apply the proposed EmJEm framework
  to learn an inference model of the articulatory parameters through iterations. 
  Each iteration involves a sampling step and a training step as illustrated in Figure \ref{Fig.whole}. 
  At the sampling step, articulatory parameters are sampled and fed to the TRM model to synthesize speech; 
  At the training step, the inference model is trained on the generated paired data in a supervised manner. 
  Note that the inference model applied during these two steps is exactly the same. 

  The model takes mel-spectrograms as input and outputs both control-rate and utterance-rate parameters, 
  with an architecture illustrated in Figure \ref{Fig.architecture}. 
  A U-Net \citep{ronneberger2015u} like CNN structure is used to obtain features with different time and frequency scales. 
  Kernel sizes of all convolutional layers are set to 3 and strides are specified to conserve sequence lengths. 
  A layer of bidirectional LSTM (BLSTM) \citep{graves2005framewise} with hidden size 128 is stacked on top of the convolutional layers. 
  The forward and backward outputs of BLSTM at every time stamp are concatenated and mapped to 16-dimensional control-rate parameters through a linear layer, 
  while the last cell states of BLSTM are concatenated and then mapped to 13-dimensional 
  utterance-rate parameters (the other 13 dimensions are fixed during training, details are given in the supplementary material) through another linear layer. 
  ReLU activation is used for all convolutional layers while tanh activation is used for the linear layers.
  Both the loss of utterance-rate parameters $L_u$ and the loss of control-rate parameters $L_c$ are mean 
  square errors, and our optimization objective is to minimize $L_u + \lambda L_c$, where $\lambda$ is a superparameter.

\subsection{Training Details}
\label{exp.setup}
  We use 80-dimensional log magnitude mel-spectrogram with 50 ms frame length and 12.5 ms frame shift, 
  following the practice in \citep{wang2017tacotron}. 
  The inferred articulatory parameters are rescaled before fed to the TRM model, 
  and the control-rate parameters are interpolated to 250 Hz to meet the requirement of the TRM model.
  Sample number per datapoint $L$ is set to $1$, so that the deterministic output of the network can be directly viewed as sampling results.
  We apply Algorithm 2 to update the training set with a data drop rate $\gamma = 0.33$.

  The neural network is randomly initialized and trained with the loss weight $\lambda = 0.001$. 
  We use the Adam optimizer with learning rate of $5\times10^{-4}$, with $\beta_1=0.5$ and $\beta_2=0.999$.
  There is no need to complete optimization in each iteration, we train for 10 epochs in a single iteration.
  Batch sizes vary from different datasets.
  We train the model on a computer with 2 NVIDIA V100 GPUs and 2 Xeon E5-2690 V4 CPUs. 
\subsection{Evaluation Metrics}
  We aim to train a model which can infer articulatory parameters from a reference utterance, 
  so as to synthesize an utterance similar to the reference one. 
  We measure the similarity between the synthetic and reference utterances by 
  the signal to noise ratio (SNR) of the synthetic mel-spectrogram $\widehat{Y}$, the higher the SNR, the more similar the synthetic and reference utterances are. The SNR was expressed as:
  \begin{equation}
  \label{eq.sent_snr}
    SNR(Y, \widehat{Y}) = 10log_{10}\frac{\sum_{i=1}^{N}\sum_{j=0}^{M}Y_{i,j}^{2}}{\sum_{i=1}^N\sum_{j=0}^{M}(Y_{i,j}-\widehat{Y}_{i,j})^2}
  \end{equation}
  where N denotes the frame number; M denotes the number of bins;
  $Y_{i,j}$ and $\widehat{Y}_{i,j}$ are the jth bin of the ith frame of the reference and synthetic mel-spectrograms, respectively.
  To check the quality of the synthetic mel-spectrogram at different time stamps and different
  frequencies, the SNR at each time-frequency point can be used, which can be expressed as:
  \begin{equation}
  \label{eq.local_snr}
    SNR(Y_{i,j}, \widehat{Y}_{i,j})=20log_{10}\frac{|Y_{i,j}|}{|Y_{i,j}-\widehat{Y}_{i,j}|}
  \end{equation}
  For clarity, we refer to $SNR(Y, \widehat{Y})$ as sentence SNR and $SNR(Y_{i,j}, \widehat{Y}_{i,j})$ as local SNR.

\section{Evaluations}
  In this section, we conduct experiments on single-speaker and multi-speaker datasets to demonstrate the 
  convergence property of EmJEm and report model performance with qualitative and quantitative results. We then
  show that trained models can generalize well to datasets of unseen speakers or even a new language, 
  and the model performance can be further improved through self-adaptation.
\subsection{Single-speaker Evaluation}
  \paragraph{Dataset and Setup} We conduct experiments on the LJSpeech dataset \citep{ljspeech17}, which contains 13,100 English audio clips of a single female speaker, with a total audio length of approximate 24 hours.  
  We randomly choose 5400 clips for training, 540 clips for validation and 540 clips for testing.
  The experimental setup follows Section \ref{exp.setup} with a batch size of 180 clips.
  \begin{figure}[H]
    \centering  
    \subfigure[Learning on single-speaker dataset (LJSpeech)]{
    \label{Fig.loss_snr}
    \includegraphics[width=0.506\textwidth]{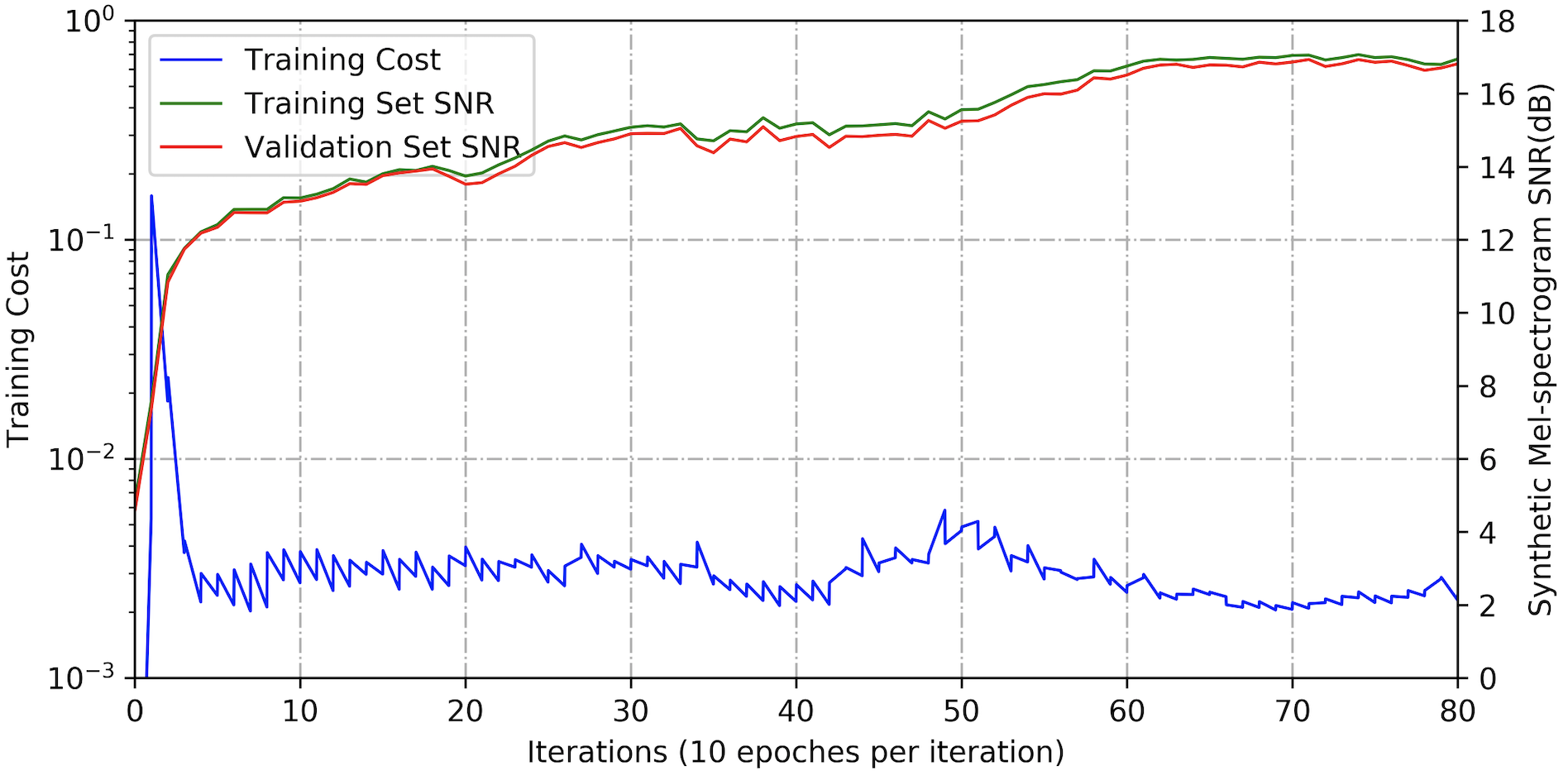}}
    \subfigure[Learning on multi-speaker dataset (ARCTIC)]{
    \label{Fig.cmu7}
    \includegraphics[width=0.472\textwidth]{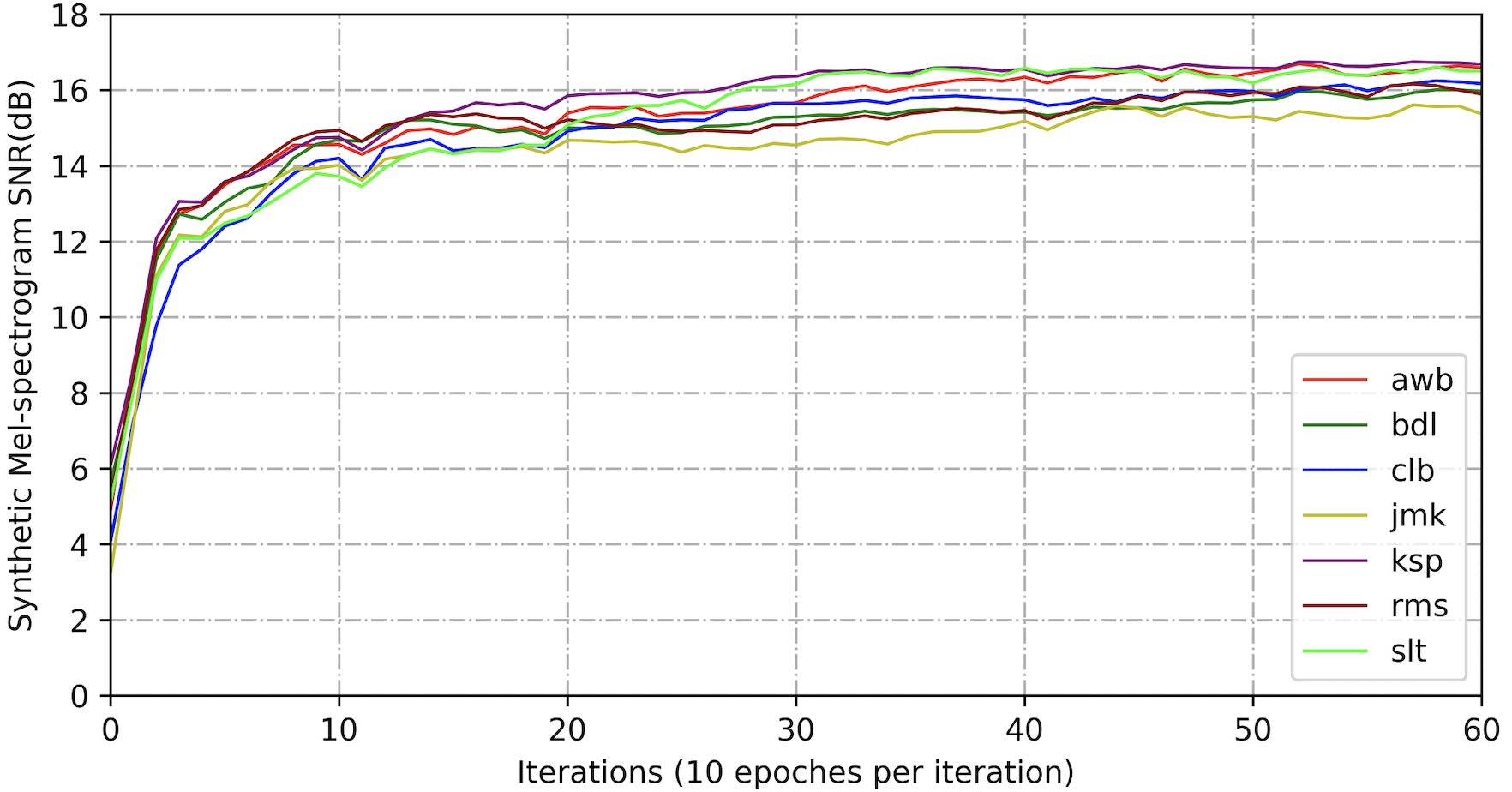}}
    \caption{Learning procedure. 
      (a) Training cost and the sentence SNR when trained on the LJSpeech dataset;
      (b) The sentence SNR on validation sets of different speakers (represented by different colors) when trained on the ARCTIC dataset.}
    \label{Fig.snr_res}
  \end{figure}
  \paragraph{Results} 
  The convergence process is shown in Figure \ref{Fig.loss_snr}.
  While the training cost reduced from $1\times10^{-1}$ to around $2\times10^{-3}$ in general, 
  there is always a cost increase at the beginning of each iteration and the training cost seems to oscillate through iterations.
  Both of the results are due to the fact that we update the training set $\Phi$ with newly sampled data at the beginning of each iteration.
  However, rather than to minimize the training cost, our purpose is indeed to improve 
  the similarity between the synthetic and reference utterances. 
  Therefore, at the beginning of each iteration, we directly evaluate the model performance on the training and validation set 
  by (i) apply the inference model to extract articulatory parameters; (ii) synthesize utterances; (iii) calculate 
  the sentence SNR of the synthetic utterances following the Equation (\ref{eq.sent_snr}).
  As illustrated in Figure \ref{Fig.loss_snr}, sentence SNR increases steadily with iterations, and stablizes after around 60 iterations. 
  Each iteration takes around 15 minutes to complete.
  We call this trained model as the LJ-model.

  We evaluate the trained model on the test set and get a mean sentence SNR of $16.81\pm0.76$dB. 
  A pair of reference and synthetic mel-spectrograms from the test set are shown in Figure \ref{Fig.spec_compare}.
  We also attach several audio samples in the supplementary material for reference.
  As demonstrated by the samples, the LJ-model successfully infers the underlying articulatory parameters of the reference utterances.
  We also calculate the local SNR of the test set as defined in Equation (\ref{eq.local_snr}) and do statistics for every mel-bin.
  As illustrated in Figure \ref{Fig.bin_snr}, the reference and the synthetic spectrograms have high similarity in details, with 
  the medians of local SNR above 20dB for most bins, while the first low frequency bins have a lower local SNR, 
  due to the fact that the energy of those bins is rather low. 
  \begin{figure}[H]
    \centering  
    \subfigure[Mel-spectrograms comparison]{
    \label{Fig.spec_compare}
    \includegraphics[width=0.39\textwidth]{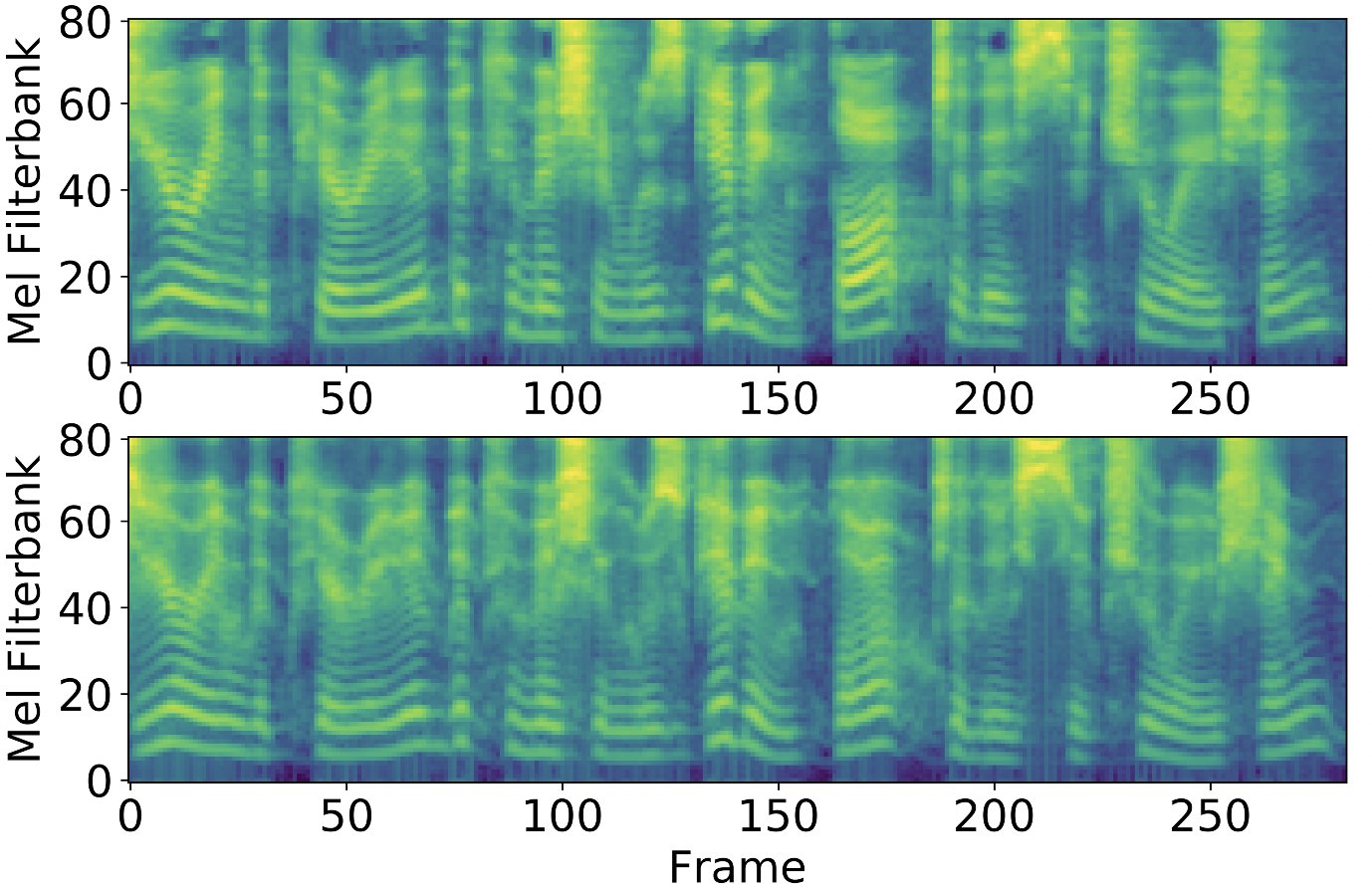}}
    \subfigure[Boxplots of local SNR for different mel-bins]{
    \label{Fig.bin_snr}
    \includegraphics[width=0.59\textwidth]{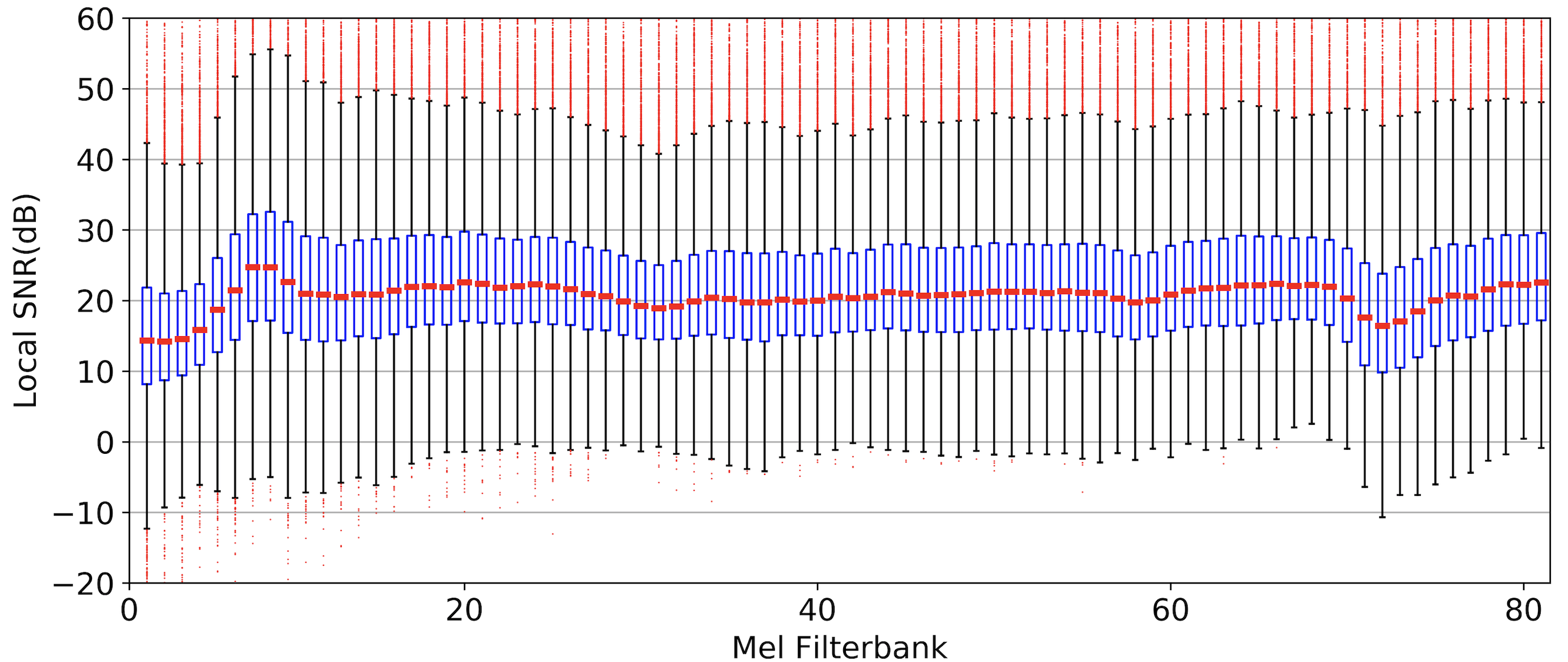}}
    \caption{Test results of the LJ-model. 
            (a) A reference (top) and the corresponding synthetic (bottom) mel-spectrograms, 
            the corresponding text is "During the period the Commission was giving thought to this situation"; 
            (b) SNR are truncated to [-20, 60], middle bars represent medians, boxes represent the interquartile 
            range(IQR; 25th-75th percentile), whiskers extend to 1.5IQR outside the IQR, and outliers are represented as single points.}
    \label{Fig.LJ_res}
  \end{figure}
\subsection{Multi-speaker Evaluation}
  \paragraph{Dataset and Setup} We conduct multi-speaker experiments on the ARCTIC dataset \citep{kominek2003cmu} with a total audio length 
  of approximate 7 hours. The dataset contains utterances of 7 speakers (5 males and 2 females), with around 1150 clips
  for each speaker. We first randomly split the utterances of each speaker into training, validation and testing sets with a ratio of 8:1:1, 
  then merge the training sets of different speakers for training, while keeping track of each speaker for validation and testing.
  The experimental setup follows Section \ref{exp.setup} with a batch size of 300 clips.
  \paragraph{Results} The sentence SNR curves on validation sets for all speakers are shown in Figure \ref{Fig.cmu7}.
  As can be seen, though there is performance difference between speakers, quality of synthetic utterances 
  improve steadily for all speakers in general. After 60 iterations, the averaged sentence SNR on the merged testing set is $16.11\pm0.91$dB,  
  and the trained model is called as the ARCTIC-model.

\subsection{Generalizing and Adapting to Unseen Speakers and New Languages}
  \label{adaptation}
\begin{figure}
    \centering  
    \subfigure[Generalize to unseen English speakers]{
    \label{Fig.multiple_test}
    \includegraphics[width=0.40\textwidth]{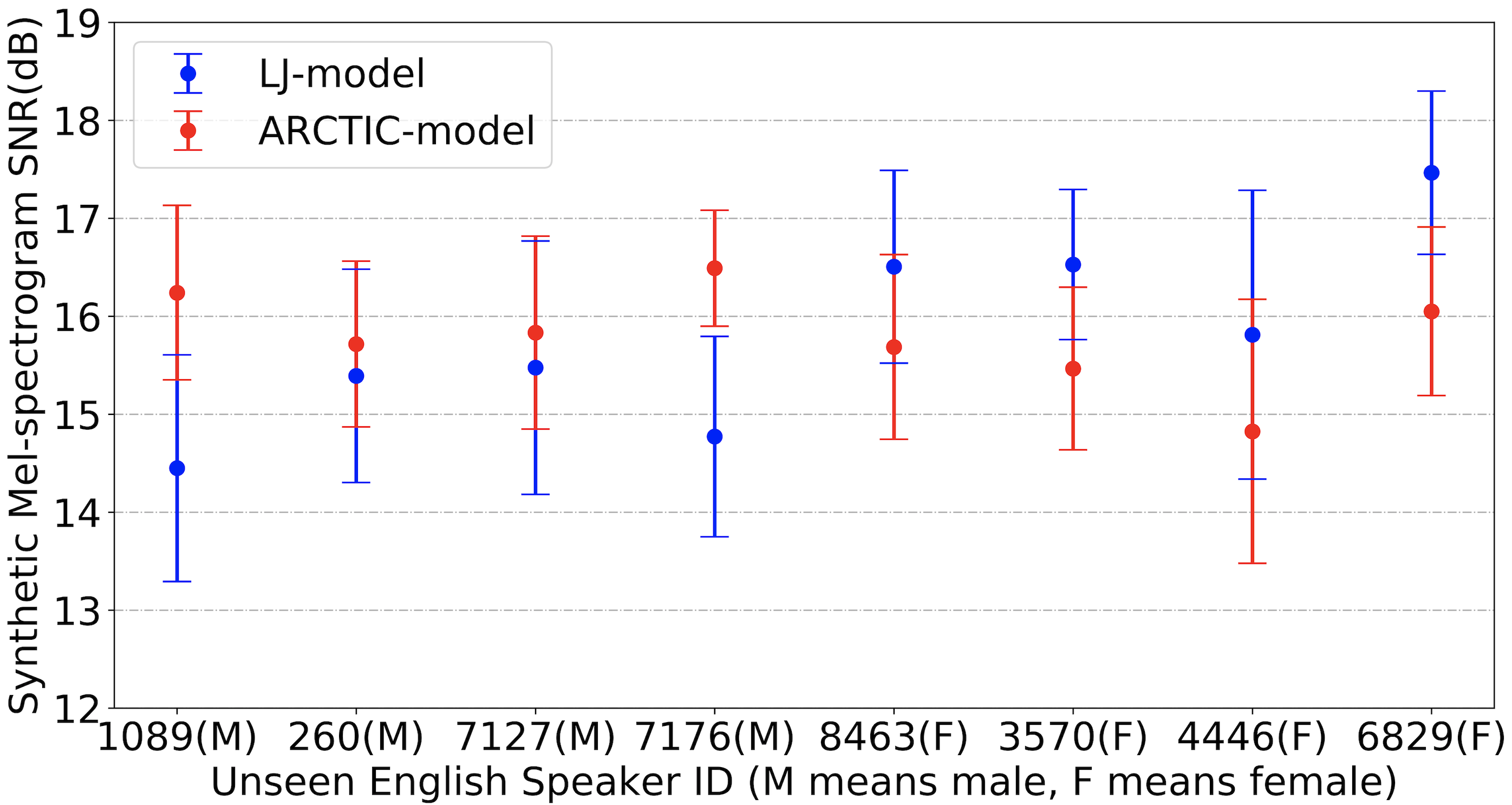}}
    \subfigure[Unseen speaker adaptation]{
    \label{Fig.1089_adaptation}
    \includegraphics[width=0.27\textwidth]{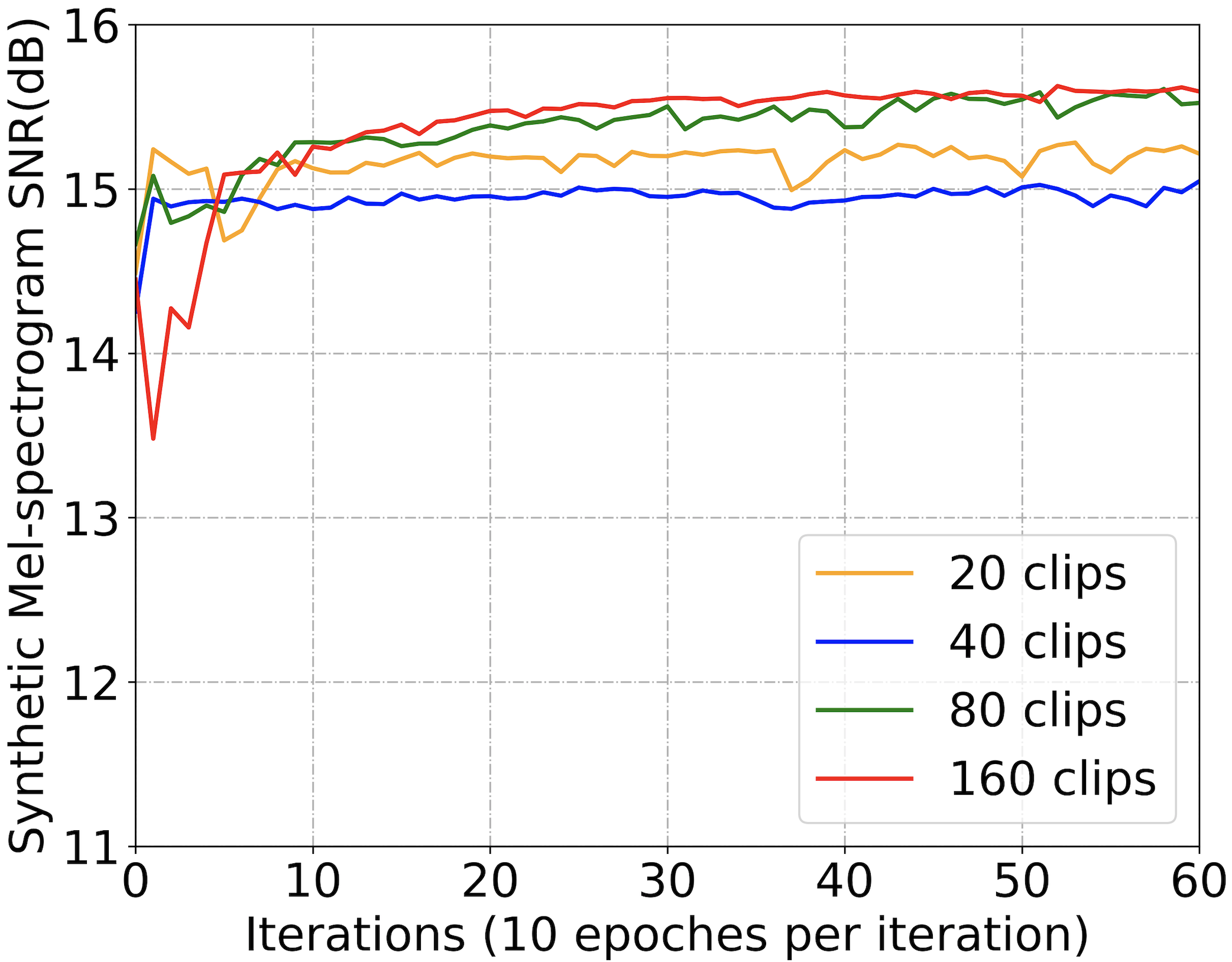}}
    \subfigure[Chinese adaptation]{
    \label{Fig.Chinese_adaptation}
    \includegraphics[width=0.27\textwidth]{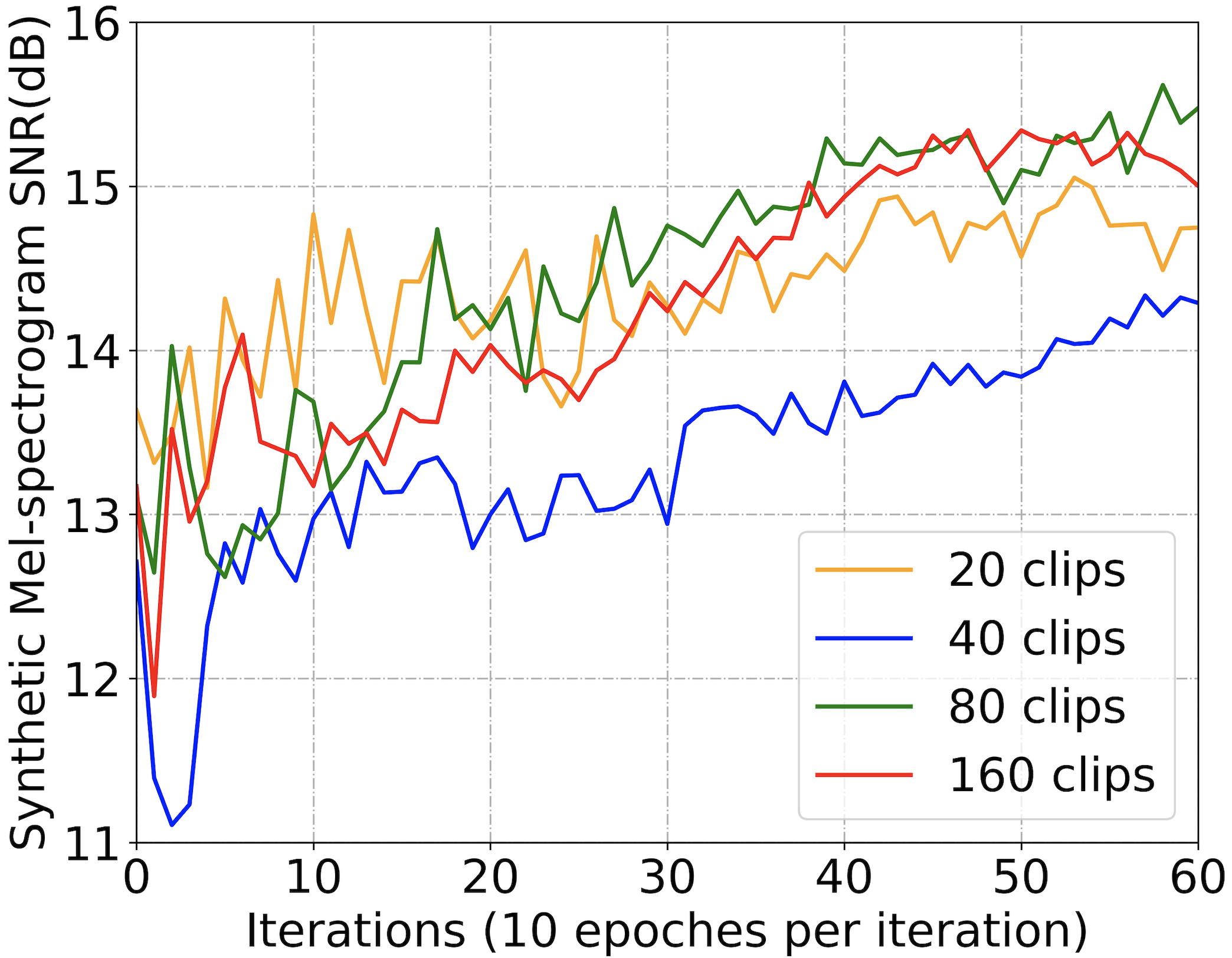}}
    \caption{Generalization and adaptation of models. (a) Test the LJ-model 
    and the ARCTIC-model on unseen English speakers (each with 160 testing clips); 
    (b) Adapt the LJ-model to an unseen English speaker (ID 1089); 
    (c) Adapt the LJ-model to a Chinese speaker.}
    \label{Fig.Generation_adaptation}
  \end{figure}

  \paragraph{Dataset} We construct test sets of unseen English speakers from the test-clean set of
  the LibriTTS Corpus \citep{zen2019libritts}. To ensure enough clips for adaptation experiments, we filter out those speakers 
  with less than 160 clips, resulting in 13 speakers (4 males and 9 females). We then randomly choose 4 female speakers along 
  with all the 4 male speakers to build separated test sets with 160 clips for each speaker.
  We construct a test set of Mandarin Chinese by randomly picked 160 clips from the Chinese set of the
  Css10 Corpus \citep{park2019css10} which is constructed from two audiobooks read by a female.
  \paragraph{Results} We evaluate the trained LJ-model and ARCTIC-model for all the 8 unseen English speakers.
  As shown in Figure \ref{Fig.multiple_test}, both the models achieve a relatively high sentence SNR (above 15dB) for most unseen speakers. 
  Meanwhile, the LJ-model does not generalize as well as the ARCTIC-model to male speakers, 
  but outperforms the ARCTIC-model on the test sets of female speakers. These differences are consistent with their training data.
  We then evaluate the LJ-model, which is trained on an English dataset, on the Chinese test set, 
  resulting in a sentence SNR of $13.17\pm1.81$dB.
  The result reflects that the LJ-model can generalize to Chinese well, though with some performance loss.
  
  Furthermore, the self-supervised nature of EmJEm makes it possible for a trained model to adapt itself
  to obtain a better estimation of underlying articulatory parameters when met unseen utterances.
  We conduct adaptation experiments with the LJ-model on the test set of Speaker 1089
  (i.e. the speaker with ID 1089, on which the LJ-model has poorest performance) and on the Chinese test set.
  To explore how few unseen utterances are enough for adaptation, we randomly pick 20, 40, 80, 160 clips from 
  the test sets and finetune the LJ-model on the picked clips.

  As shown in Figure \ref{Fig.1089_adaptation} and Figure \ref{Fig.Chinese_adaptation}, sentence SNR on the picked clips
  continuously improves as adaptation proceeds,
  and generally, adaptation stability and model performance benefit from more adaptation samples. 
  After adaptation, sentence SNR improves from $14.45\pm1.16$dB to $15.59\pm0.82$dB on the test set of Speaker 1089 and 
  from $13.17\pm1.81$dB to $15.00\pm1.77$dB on the Chinese test set, which means adaptation is an effective way 
  to transfer trained models to unseen data.

\section{Conclusion and Future Work} 
  We have proposed a novel approach EmJEm to solve inverse problems in a self-supervised manner,
  so as to obtain a physically meaningful representation of the observed data. 
  By integrating a physical forward process, the proposed approach works in an analysis-by-synthesis procedure 
  by iteratively sampling and training, which can be viewed as a form of embodied learning.
  We verify the proposed method by tackling the acoustic-to-articulatory inversion problem. 
  Given an articulatory synthesizer and reference utterances, 
  the model learns from scratch to extract articulatory parameters to synthesize speech very close to the reference utterances.
  Besides, our experiments demonstrate that a trained model can be transferred to unseen speakers or even a new language through self-adaptation. 

  Articulatory kinematic information can benefit pronunciation teaching, phonology research, machine speech 
  recognition and synthesis. To be applied in such fields, properties of the extracted 
  articulatory information need to be further analysed in future.
  
  The proposed EmJEm framework can be applied to solve inverse problems in those fields with forward processes, such as 
  simulated or physical robots or manipulators, especially when 
  input parameters are non-trivial to sample and the forward processes are non-differentiable.
  
  Last but not least, due to motor equivalence phenomena \citep{perrier201511}, acoustic-to-articulatory 
  inversion suffers from the problem of non-uniqueness, which is also common among other inverse problems. 
  Attention should be paid to such aspects when applying EmJEm to solve inverse problems.

\section*{Broader Impact}
  a) If the proposed self-supervised method is adopted, the resources that would otherwise be used to obtain labelled data can be saved.
  When used to extract articulatory information as in our experiments, the method can benefit language 
  cultural preservation and second language learning.
  Specifically, the method can benefit phonology research so as to better protect language culture, 
  especially for dialects and minority languages; 
  while in education, the method can promote pronunciation teaching.
  
  b) As far as we have concerned, nobody will be put at disadvantage from this research.

  c) The proposed method can check whether success or failure when running, thus, no external consequences will occur even the system fails.

  d) The proposed method does not have the problem of leveraging the biases in the data.
  
\begin{ack}
The work was supported in part by the National Natural Science Foundation of China (No. 11590773) 
and the National Social Science Foundation of China (No. 15ZDB111).
We also acknowledge the High-Performance Computing Platform of Peking University for providing computational resources.
\end{ack}

\small
\bibliographystyle{plainnat}
\bibliography{mycite}

\end{document}